\documentclass[12pt]{iopart}
\usepackage{graphicx}
\usepackage{caption}%
\usepackage{color}
\usepackage{epstopdf}
\usepackage{threeparttable}
\usepackage{multirow}

\begin{document}
	
	\title[]{Possible origin for the similar phase transitions in $k$-core and interdependent networks}
	
	\author{ Shengling Gao$^{1,2}$, Leyang Xue$^{2,4}$,  Bnaya Gross$^{2}$,  Zhikun She$^{1}$,  Daqing Li$^{3}$, Shlomo Havlin$^{2}$}
	
	\address{$^{1}$School of Mathematical Sciences, Beihang University, Beijing, 100191, China.}
	
	\address{$^{2}$Department of Physics, Bar-Ilan University, Ramat-Gan, 52900, Israel}
	
	\address{$^{3}$School of Reliability and Systems Engineering, Beihang University, Beijing, 100191, China.}
	
	\address{$^{4}$International Academic Center of Complex Systems, Beijing Normal University, Zhuhai, 519087, China}
	
	\ead{havlins@gmail.com}
	
	\vspace{10pt}
	\begin{indented}
		\item[]4/8/2023
	\end{indented}
	
	\begin{abstract}
		The models of $k$-core percolation and interdependent networks (IN) have been extensively studied in their respective fields.
		A recent study has revealed that they share several common critical exponents. However, several newly discovered exponents in IN have not been explored in $k$-core percolation, and the origin of the similarity still remains unclear.
		Here,  we investigate k-core percolation in random networks. We find that for k-core percolation,
		the fractality of the giant component fluctuations is manifested by a fractal fluctuation dimension, $\widetilde d_f = 3/4$, 
		within a correlation \emph{size} $N'$ that scales as $N' \propto (p-p_c)^{-\widetilde\nu}$, with $\widetilde\nu = 2$, same as found in IN. Indeed, here, $\widetilde\nu \equiv d\cdot \nu'$ and $\widetilde{d}_f \equiv   d'_f/d$, where $\nu'$ and $d'_f$ are respectively the same as the correlation  \emph{length} exponent and the fractal fluctuation dimension observed in $d$-dimensional IN spatial networks.
		These two new exponents found here for $k$-core percolation demonstrate the same scaling behaviors as found for IN with the same critical exponents,
		reinforcing the similarity between the two models.
		Furthermore, we suggest that these two models are similar since both have two types of interactions: short-range (SR) connectivity  and long-range (LR) influences.  In IN the LR are the influences of dependency links while in k-core we find here that for $k=1$ and $k=2$ the influences are short range while for $k\geq3$ the influence is long range.
		In addition, analytical arguments for a universal hyper-scaling relation for the fractal fluctuation dimension of the $k$-core giant component and for IN as well as for any mixed-order transition are established.
		Our analysis enhances the comprehension of k-core percolation and supports the generalization of the concept of  fractal fluctuations in mixed-order phase transitions.
	\end{abstract}
	
	%
	
	\section{Introduction}\label{s1}
	
	The $k$-core model has been studied extensively since its introduction in 1960 to explore the question of what is the minimal number of colors required for covering graphs \cite{erdHos1960evolution,erdos1963structure,erdHos1966chromatic,matula1968min,szekeres1968inequality}.
	The model has a significant impact on the field of network science, and has been used to describe real-world systems such as social networks~\cite{bhawalkar2015preventing}, 
	the Internet~\cite{carmi2007model}, ecological networks~\cite{morone2019k}, transportation networks~\cite{wang2014evolution}, influential spreaders~\cite{kitsak2010identification},	etc. 
	The k-core model can facilitate a better understanding of the organizational structure and behavioral characteristics of systems, by identifying the hierarchy of node connection patterns as well as the most densely connected subgraphs within a network called nucleus~\cite{wu2023rigorous,kong2019k}. The model can also provide insights in revealing the robustness and vulnerability of networks under attacks or failures~\cite{dorogovtsev2006k,goltsev2006k,yuan2016k}.
	
	The $k$-core percolation, recognized as an important application of the $k$-core model, primarily investigates the phase transition phenomena and critical behavior in network  breakdowns. This process is accomplished by theory and simulations of the failure and disintegration of a network. One can initiate the failure through random removal of nodes, and then iteratively prune the graph by eliminating all nodes with a degree below $k$ until
	the largest $k$-core remains.
	A $k$-core is the giant component of the original graph, where every node in the giant component has at least a degree  $k$.  
	The phase transition phenomenon in $k$-core percolation is typically characterized by the appearance and disappearance of a giant $k$-core component of the order of the size of the original network. 
	Research studies have shown that $k$-core percolation undergoes a mixed-order phase transition for $k\geq3$, 
	characterized by an abrupt jump in the order parameter (giant component) that resembles a first-order phase transition. However, it also exhibits scaling behaviors at and near the critical threshold, as typically observed in second-order phase transitions~\cite{dorogovtsev2006k,chalupa1979bootstrap,schwarz2006onset}. Further, considering $k$-core percolation for $k\geq3$ in random networks, studies have revealed  several critical exponents near the mixed-order phase transition threshold~\cite{lee2016critical}, 
	containing the critical exponents $\beta_S = 1/2$,  $\gamma_S=1$, etc., 
	defined by
	\begin{eqnarray}
		&S(z)-S_c \propto (z-z_c)^{\beta_S} 
		\label{beta_S},\\
		& \chi_S=N\left( \left\langle S^2 \right\rangle -\left\langle S \right\rangle^2\right)  \propto (z-z_c)^{-\gamma_S} 
		\label{gamma_S},
	\end{eqnarray}
	where $S(z)$ and $S_c$ denote the fraction of the $k$-core giant component of a network with size $N$ at mean degree $z$ and at critical mean degree $z_c$, respectively;
	$\chi_S$ denotes the fluctuations of $S$; 
	$\left\langle  X \right\rangle$ represents the mean value of $X$.
	These exponents yield insights into the nature of critical phenomena that occur near phase transitions of the system, such as the scaling behavior near the critical transition. 
	
	In recent years, a model called interdependent networks (IN) has been developed and studied by Buldyrev et~\cite{buldyrev2010catastrophic}. 
	This model has gained increasing attention~\cite{gao2012networks,parshani2010interdependent,huang2011robustness,boccaletti2014structure,lee2017universal}, 
	due to the growing  interdependence between various systems in our modern world~\cite{yang2017small,li2015percolation}.
	The interdependent networks system is defined as  two or more  networks where pairs of nodes in different  networks depend for functioning on each other. 
	Thus, the failure of a node in one network triggers the failure of its dependent nodes in another network, as well as the failures of those nodes that are connected to the network via the failed nodes. Note that dependency links may also exist within a single network, yielding similar results \cite{parshani2011critical,bashan2011percolation}.
	
	By studying interdependent networks, 
	researchers can gain insights into the robustness and resilience of such interacting macroscopic systems, 
	as well as exploring the mechanisms that drive cascading failures and collapse transition. 
	In particular, studying the critical behavior of interdependent networks can help us in providing   insights into the mechanisms  yielding the critical phenomena and help to design more resilient interdependent systems.
	Noteworthy,
	research studies have shown that interdependent networks (IN) also undergo, 
	like k-core~\cite{dorogovtsev2006k,goltsev2006k,lee2016critical}, mixed-order phase transition~\cite{parshani2010interdependent,lee2016hybrid,gao2011robustness,hu2011percolation}. 
	Moreover, 
	in percolation of interdependent networks, 
	the critical exponents $\beta$ and $\gamma$ also have respectively values equal to $1/2$ and $1$ like in k-core~\cite{dorogovtsev2006k,lee2016critical,zhu2017revealing,li2021percolation}, near the mixed-order phase transition point~\cite{parshani2010interdependent,lee2016hybrid,dong2018resilience}.
	
	Zhou et al.~\cite{zhou2014simultaneous} studied the dynamics of cascading failures in IN and found that
	the mean value of the plateau time, $\tau$ (Fig.~\ref{fig4}(a)), that is the number of iterations (time) until the system fully collapses at $p_c$ scales as $\left\langle \tau_c\right\rangle \propto N^{1/3}$ and $\left\langle \tau_c\right\rangle \propto(p_c-p)^{-1/2}$. 
	Very recently, 
	Gross et al.~\cite{gross2022fractal} have identified in IN critical characteristics of the order parameter fluctuations near the threshold of a mixed-order phase transition.
	They find that
	the fluctuations of the order parameter exhibit a fractal fluctuation dimension, $d'_f$, for length scales up to the correlation \emph{length}. 
	This is analogous to continuous second-order transitions, where near criticality the order parameter itself is a fractal. 
	Moreover, 
	Gross et al.~\cite{gross2022fractal}  found that for IN the hyperscaling relation between $d'_f$, 
	correlation \emph{length} exponent $\nu'$ and $\beta$ is valid, i.e.,
	\begin{eqnarray}
		d'_f=d-\beta/\nu', \label{hyper}  
	\end{eqnarray}
	for any $d$.
	Note that these two new exponents (i.e., $d'_f$ and $\nu'$) also exist in  interdependent random networks \cite{gross2022fractal}, like Erd\H{o}s-R\'enyi (ER) networks~\cite{bollobas1998random}. 
	
	These two models, $k$-core and IN, have flourished in their respective fields. 
	Lee et al. \cite{lee2016critical} have identified several shared behaviors and features between these two models.
	We claim here that in $k$-core percolation, like in IN,
	nodes fail due to two types of interactions that have different length scales. 
	A node can fail because it does not belong to the giant component or because its degree is below k.
	In interdependent network percolation, 
	nodes are also removed based on becoming isolated from their own network or due to their dependencies on failing nodes in other networks or in the same network \cite{parshani2011critical}. 
	Thus, in both models, there exist two different types of interactions.
	In addition, 
	percolation in both models exhibits mixed-order phase transition, with the same exponents $\beta=1/2$ ~as well as the same scaling of the plateau times $\left\langle \tau_c\right\rangle \propto N^{1/3}$ \cite{lee2016critical,lee2016hybrid,zhou2014simultaneous} near criticality.
	Lee et al. \cite{lee2017universal} revealed that the universal mechanism for mix-order phase transitions may be related to long-range loops \cite{lee2017universal}.
	The two models indeed exhibit some similarities, but it remains unclear whether the newly discovered exponents for correlation \emph{length} and fractal fluctuations found for interdependent networks (IN)~\cite{gross2022fractal} also exist in $k$-core percolation with the same values. Furthermore, the origin of the similarity in the phase transitions of the two models is also unclear.
	
	Thus, in the present manuscript, we explore these two questions,  the similarity between the models and the mechanisms behind, by investigating $k$-core percolation in ER random networks. Specifically, 
	using the node occupancy $p$ as the controlled variable and the $k$-core size as the order parameter, 
	we examine, like in \cite{gross2022fractal}, the  fluctuations 
	of the critical threshold and of the order parameter, 
	as a function of $N$ and $p-p_c$. 
	Interestingly, we find here that $k$-core percolation also exhibits a fractal fluctuation dimension $\widetilde{d}_f\equiv d_f'/d=3/4$ for network sizes up to the correlation \emph{size} $N'$, where $N'\propto(p-p_c)^{-\widetilde\nu}$ with $\widetilde\nu\equiv d\nu'=2$. 
	Note that $\widetilde\nu$ ($\widetilde{d}_f$) is obtained by the scaling of the fluctuations of $p_c$ ($M_c$) with respect to the network size $N$. 
	Table \ref{tab1} summarizes the critical exponents of $k$-core percolation and IN percolation in random networks.
	The identical critical exponents shared by both models, as presented in Table \ref{tab1} provide strong support for the similarity between the $k$-core and IN models. 
	It is very plausible that the origin for the similarity between both systems (and probably in many other systems), lies in the similar fundamental mechanisms of these models. Specifically, the presence of two types of  interactions: one is short range that is a node fail when becoming disconnected from the network and the other is long range influence by having a degree below $k$ (in $k$-core)  or not being supported via a dependency link (in IN). As shown below, the distribution of the distances caused
	by a failure in different $k$-cores provides strong support for this hypothesis.
	Our study contributes to a deeper understanding of the mechanisms of the  $k$-core percolation, and also supports the generalization of the fractal fluctuations phenomena as well as the diverging correlation \emph{length} and their meaning in general mixed-order phase transitions.
	
	\section{Possible origin for the similarity of phase transition  between $k$-core model and interdependent networks}\label{s2}	
	In this Section, we will explore the critical behavior of  $k$-core percolation 
	in Section \ref{s2.1} and of IN in Section \ref{s2.2}, both for ER random networks of size $N$ with a mean degree $z$. We will also further try to uncover the origin of the similarity between both systems.
	\subsection{Short-range and long-range interactions  in k-core percolation}\label{s2.1}	
\begin{figure}[htp]
		\centering
		\includegraphics[width=\columnwidth]{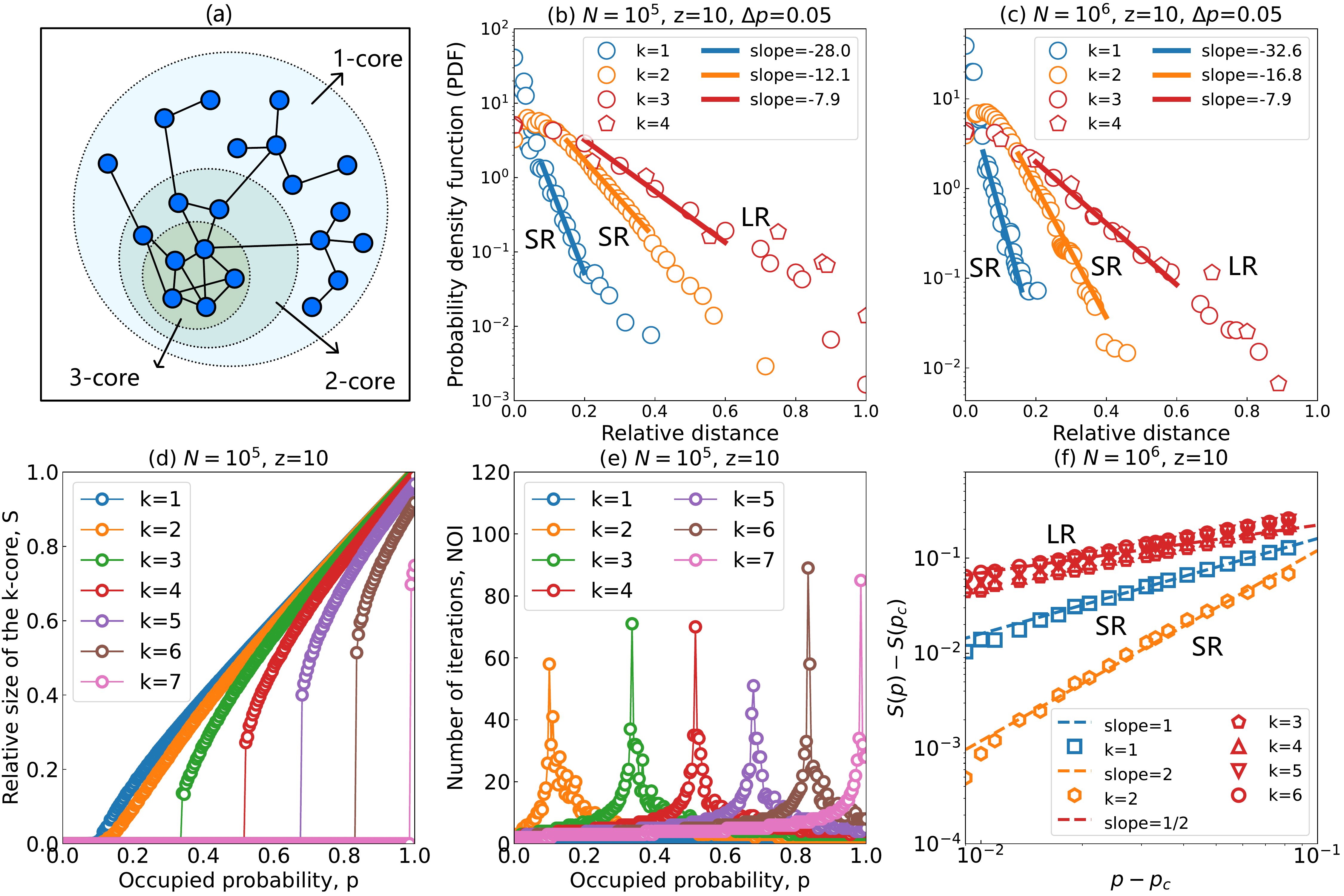}
		\caption{Results for k-core percolation.  
			(a) Demonstration of the $k$-core giant component. The subgraphs surrounded by circles of different colors represent distinct cores.  
			(b) Probability density distribution (PDF) of the relative distance of failures caused by the removal of single nodes in the $k$-core giant component. The $k$-core giant component is obtained from the original ER network of network size $N=10^5$ and mean degree $z=10$ at node occupied fraction $p$ near critical threshold $p_c$, satisfying $\Delta p=p-p_c=0.05$. (c) Similar to (b) for the case $N=10^6$. We can clearly see in (b) and (c) for $k=1$ and $k=2$, only short-range (SR) interactions exist. This is since in these cases, the slope increases when $N$ increases. However, for $k\geq3$, the slope remains the same, meaning that long-range (LR) interactions exist.
			(d) The fraction of $k$-core giant component, $S$ (the order parameter), as a function of the occupied fraction $p$ of nodes for different $k$ in ER network of size $N=10^5$ and mean degree $z=10$.  
			(e) Number of iterations (NOI) to reach the steady state  for different $p$.  
			The critical threshold of the abrupt first order phase transition can be identified by the $p$ at which NOI reaches maximum (as found for IN~\cite{bashan2013extreme}). 
			(f) The scaling law $S(p)-S(p_c)\propto (p-p_c)^\beta$ exists near the  threshold $p_c$ for different $k$. 
			We have continuous transition with $\beta=1$ for $k=1$ (SR interactions), and $\beta=2$ for $k=2$ (SR interactions, but a slightly longer than $k=1$) and $\beta=1/2$ for $k\geq3$ (LR interactions), as found also in~\cite{dorogovtsev2006k,lee2016critical}. Here we use a single  realization of ER network of size $N=10^6$ and mean degree $z=10$.}
		\label{fig1_kcore}
	\end{figure}

	The $k$-core is the maximal  subgraph of the original graph, where every node in the $k$-core giant component has at least a degree $k$, as shown in Fig.~\ref{fig1_kcore}(a).
	To obtain the $k$-core giant component of the original network, 
	we first remove randomly a fraction of $1-p$ nodes from the original network.
	We now iteratively remove nodes, where at each iteration all nodes with a degree below $k$ are removed until there are no such nodes. 
	If a $k$-core of the order of $N$ nodes remains, we are above the transition $p_c$ and if such a giant component does not exist, we are below $p_c$. 
	
	To test the range of influence,  we randomly remove  a node from the $k$-core giant component, where the giant component is obtained at node occupation rate $p(k)$ where  $p(k)$ is close and above the critical threshold $p_c(k)$ such that $\Delta p=p(k)-p_c(k)= 0.05$. 
	This removal will cause some nodes in the k-core giant component to have a degree below $k$, thereby triggering the removal of further nodes. This process will generate an avalanche and we will test here, how far, that is how many layers, in the network are influenced by the removed node. By analyzing  this distribution, we expect to distinguish between short range and long range interactions. 
	We plot in Fig.~\ref{fig1_kcore}(b) and \ref{fig1_kcore}(c) the probability density distribution (PDF) of the relative maximal distance affected by the avalanche of the removed node in the $k$-core giant component. 
	The relative distance of the initially removed node is defined as the ratio between its propagation distance and its maximum potential propagation distance, ranging from 0 to 1. The propagation distance of the initially removed node represents
	the maximum shortest path between subsequently removed nodes and the initially removed
	node, while the maximum possible  propagation distance corresponds to the maximum shortest path between the nodes in the k-core giant component and the initially removed node.
	Small ratios that decrease with network size correspond to short range (SR) interactions and large ratios which do not change with network size, correspond to long range (LR) interactions. 
	We can clearly see in Fig.~\ref{fig1_kcore}(b) and \ref{fig1_kcore}(c) 
	that for $k=1$ and $k=2$, the process features SR interactions. However, for $k\geq3$, 
	all lines collapse together and behave as LR interactions. Moreover, 
	the linearity in the log-linear plot suggests that the distribution follows an exponential distribution $P(\ell)\sim e^{-\lambda \ell}$ of distances $\ell$. The slopes $\lambda$ reveal that the decay for small $k$ ($k=1$ and $k=2$) is faster than that of larger $k$ ($k\geq 3$). Note that, the decay for $k=1$ is also faster than that of $k=2$.  Importantly, note that the slopes $\lambda$ for larger $k$ ($k\geq 3$) do not change with network size and thus supporting the hypothesis that the effect is LR. In contrast, for $k$=1 and $k$=2, the slopes increase (relative distances become smaller)  with network size,  suggesting they are of SR nature. As we will see later, in both systems k-core and IN, SR interactions yield a second order phase transition while LR interactions yield a mixed order  phase transition.

	We also plot in Fig. \ref{fig1_kcore}(d)  the fraction of the $k$-core giant component, $S$, in the $k$-core model with respect to the original network of mean degree $z$ and size $N$,  
	as a function of node occupation probability, $p$, 
	for several values of $k$. 
	Two distinct types of phase transitions can be observed.
	A second-order continuous phase transition can be seen for $k=1$ and $k=2$; and a first-order abrupt transition is seen for $k\geq3$, 
	in agreement with earlier results~\cite{dorogovtsev2006k}.
	For each curve in Fig. \ref{fig1_kcore}(d) except for $k=1$, 
	we can identify the critical threshold $p_c$ by identifying the $p$-value of the maximum number of iterations (NOI). This is since we expect that NOI diverges at $p_c$ when $N$ approaches infinite.
	Indeed, the values of $p_c$ are in agreement with the theoretical values~\cite{dorogovtsev2006k}.
	Note that this method of identifying $p_c$ in $k$-core percolation is the same as that of identifying $p_c$ in interdependent networks~\cite{bashan2013extreme}. This method can help to identify $p_c$ for systems such as spatial networks where a theory for $p_c$ does not exist. 
	Moreover, as shown in Fig. \ref{fig1_kcore}(f), for the k-core model,  
	near the critical threshold, the scaling law $S(p)-S(p_c)\propto (p-p_c)^\beta$ exists, 
	where $\beta =1$ holds for $k=1$ (like in regular percolation, corresponds to SR interactions); 
	$\beta$=2 holds for $k=2$ (corresponds to SR interactions, Fig. \ref{fig1_kcore}(b) and \ref{fig1_kcore}(c), but somewhat longer than $k=1$); 
	and $\beta$=1/2 holds for all $k\geq3$ (see earlier results~\cite{dorogovtsev2006k,lee2016critical}). Later we will show that very similar results also appear in IN. 
	
	\subsection{Short-range and long-range interactions in IN percolation}\label{s2.2}	
\begin{figure}[htp]
		\centering
		\includegraphics[width=\columnwidth]{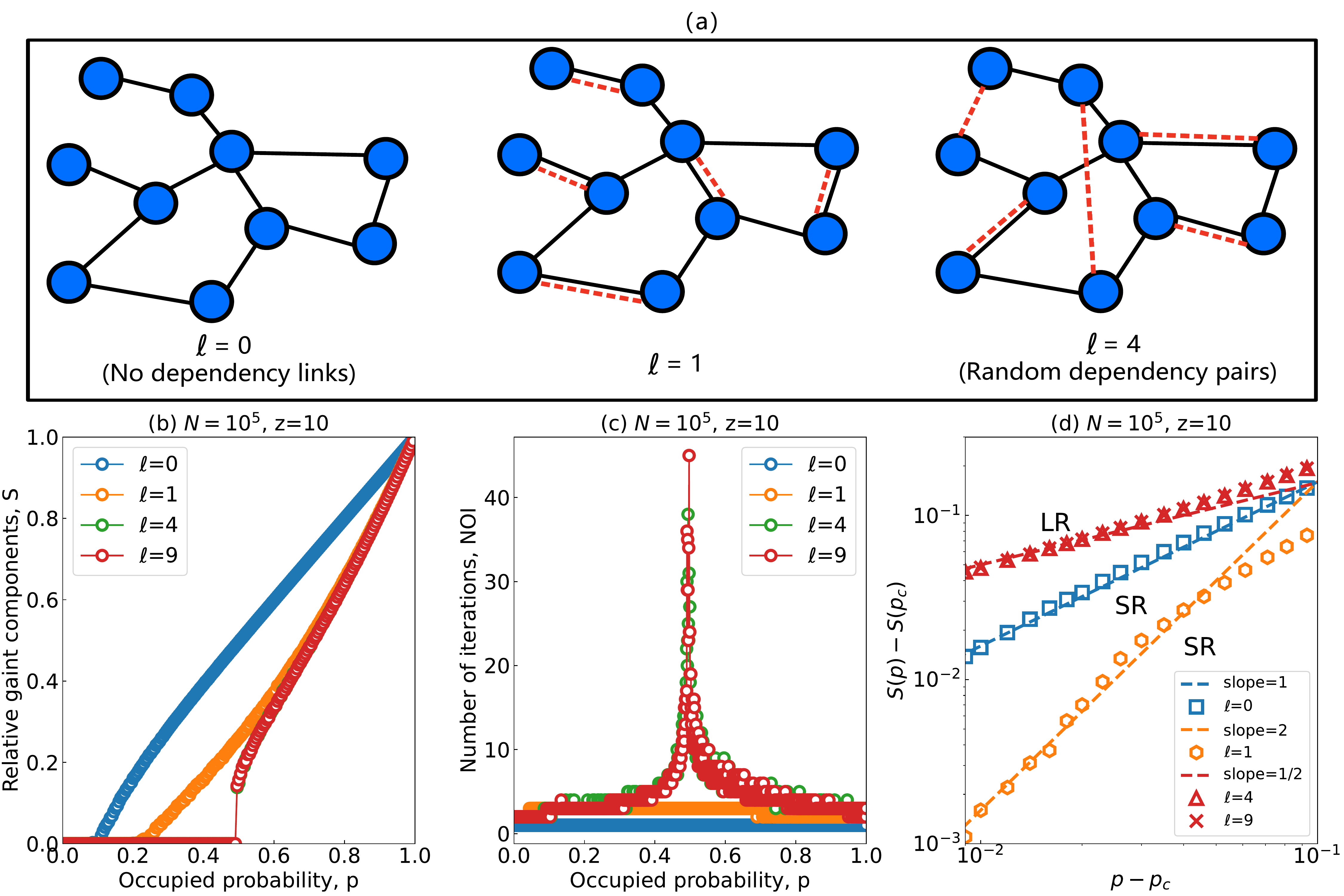}
		\caption{Results for percolation for the connectivity-dependency (CD)  model. 
			(a) Demonstration of the CD model, i.e., a single-layer network model that incorporates both connectivity links (solid black lines) and dependency links (dashed red lines), where the dependency links are subjected to a maximal shortest path length ($\ell$) constraint.  Small values of $\ell$ correspond to SR interactions, while large values of $\ell$ correspond to LR interactions. When $\ell\geq D$, where $D$ is the network diameter (in this case, $D\cong4$), the pairs of dependent nodes are chosen randomly. 		
			(b) The fraction of the giant component, $S$, as a function of the random occupied fraction $p$ of nodes for different dependency length  $\ell$ for ER random network of size $N=10^5$ and mean degree $z=10$.  
			(c)  Number of iterations (NOI) to reach the steady state  for different $p$.  
			The critical threshold of abrupt first order phase transition can be identified by the $p$ at which NOI reaches maximum (as found for $k$-core (Fig.~\ref{fig1_kcore}(e)) and for IN~\cite{bashan2013extreme}).
			(d) The scaling law $S(p)-S(p_c)\propto (p-p_c)^\beta$ also exists near the  threshold $p_c$ for different distance $\ell$. We have $\beta=1$ for $\ell=0$ (SR), and $\beta=2$ for $\ell=1$ (SR) and $\beta=1/2$~\cite{parshani2010interdependent,gross2020interconnections} for $\ell=4,9$ (LR). Note that for $\ell=D$, Parshani et al. \cite{parshani2011critical} derived an analytical  formula (see Eq.~(\ref{theo})), from which we can analytically derive the critical exponent $\beta=1/2$ (see \ref{A4} for more details), supporting our results here. Note that for obtaining  Fig.~\ref{fig1_dependency}(d) we averaged over 100 realizations of ER network of  size $N=10^5$ and mean degree $z=10$.}
		\label{fig1_dependency}
	\end{figure}

	In this Subsection, we consider interdependent networks where the dependency links are in the same network. 
	As shown in Fig.~\ref{fig1_dependency}(a), the single-layer network model incorporates both  connectivity links (solid black lines) and dependency links (dashed red lines), where the dependency links are subject to a maximal shortest path length ($\ell$) constraint. For simplicity, we denote this model as the connectivity-dependency (CD) model. Each node $i$ has one and only one dependency node $j$, with the constraint that the distance between the node $i$ and its dependency node $j$ is at most $\ell$. Small values of $\ell$ correspond to SR interactions, while large values of $\ell$ of the order of the system diameter, correspond to LR interactions.  When $\ell=0$, the single-layer network does not have dependency links, like regular percolation, and therefore $p_c=1/z$, see Fig.~\ref{fig1_dependency}(b). When $\ell\geq D$, where $D$ is the network diameter (in  Fig.~\ref{fig1_dependency}(a) case, $D\cong 4$), the pairs of dependent nodes are actually chosen randomly, like in the model of IN \cite{buldyrev2010catastrophic} or of a single network \cite{bashan2011percolation}. Note that the dependency links here are bidirectional, i.e.,  if a node $i$ fails, it will cause the dependent node $j$ to also fail, and vice versa.
	
	Now we simulate the percolation process on the CD model. First, we construct the CD model upon an ER random network of size $N$ with a mean degree $z$. In the first step, we randomly remove a fraction of $1-p$ of the nodes from the CD model. We remove the nodes iteratively while applying two processes. One is the percolation process: the failed nodes and their connectivity links are removed, causing other nodes to become disconnected from the network and fail. The other is the dependency process: the failed nodes trigger the failure of  their dependent nodes, although these dependent nodes are still connected to the network via connectivity links. The iteration terminates when no such nodes are left. 
	Similar to the k-core percolation, 	if a giant component of the order of $N$ nodes remains, we are above the transition $p_c$ and if such a giant component does not exist, we are below $p_c$.
	
	Fig. \ref{fig1_dependency}(b) depicts the fraction of the giant component, $S$ (the order parameter), in the CD model with respect to the original network with mean degree $z$ and size $N$,  
	as a function of node occupation probability, $p$, 
	for several values of $\ell$. 
	Like in $k$-core percolation (see Fig. \ref{fig1_kcore}(d)), 
	two distinct types of phase transition can be observed.
	A second-order continuous phase transition is observed in Fig. \ref{fig1_dependency}(b) for $\ell=0$ and $\ell=1$, while a first-order abrupt transition is observed for $\ell=4,9$. Notably, for $\ell=D\cong 9$, the results align with the findings in \cite{parshani2011critical}.
	For each curve in Fig. \ref{fig1_dependency}(c) except for $\ell=0,1$, 
	we can identify the critical threshold $p_c$ by determining  the $p$-value at the maximum number of iterations (NOI), like in the $k$-core model  (see Fig. \ref{fig1_kcore}(e)) and like in the IN model~\cite{bashan2013extreme}.
	Moreover, as shown in Fig. \ref{fig1_dependency}(d),  
	near the critical threshold, the scaling law $S(p)-S(p_c)\propto (p-p_c)^\beta$ exists, 
	where $\beta =1$ holds for $\ell=0$ (the limit of regular percolation); 
	$\beta$=2 is found for $\ell=1$; 
	and $\beta$=1/2 holds for $\ell=4,9$.
	
	Thus, we hypothesize  that in both models, k-core and IN,  the mechanisms behind the different phase transitions and therefore the outcome seem to be the same. In k-core percolation, the distribution of relative interaction distances  indicates that: $k=1$ and $k=2$ correspond to SR interactions, while $k\geq 3$ corresponds to LR interactions (see Fig.~\ref{fig1_kcore}(b) and (c)). Similarly, in the proposed CD model, we impose a restriction on the interaction distance $\ell$ of dependency links, where $\ell=0$ and $\ell=1$ represent SR interactions, and $\ell=4,9$ represents LR interactions. The behavior of the two models is similar for both the SR case, as well as LR case.
	Specifically, the comparison between Fig.~\ref{fig1_kcore}(f) and \ref{fig1_dependency}(d) indicates that: $\beta=1$ holds for both the CD model percolation ($\ell=0$) and the k-core percolation ($k=1$), corresponds to SR interactions; $\beta=2$ is found for both the CD model percolation ($\ell=1$) and the k-core percolation ($k=2$), also associated with SR interactions, but with slightly longer interactions compared to $\beta=1$. Furthermore, $\beta=1/2$ is found for both the CD model percolation ($\ell=4,9$) and the k-core percolation ($k\geq3$), which corresponds to LR interactions. The high consistency between the results and the critical exponents of the two models provides strong support for their similar fundamental mechanisms. Specifically, the similar behaviors observed in both models further support the hypothesis that both short-range and long-range influences are the origin of their similarity. 
	
	In the following sections, we aim to identify several new critical exponents of k-core to further support the mechanisms behind the similarity between the two models. For a summary of all known exponents and all new exponents found here, see Table \ref{tab1}.
	\begin{table}
		\caption{\label{exponents summary}Summary of critical exponents of $k$-core percolation and interdependent networks (IN) percolation of ER network.}
		\label{tab1}
		\footnotesize
		\centering
		\begin{tabular}{c|c|c|c}
			\hline
			\hline
			Scaling & Exponent & $k$-core $\left( k\geq3\right)$ & Interdependent networks ($\ell=D$)$^{a}$ \\  \hline 
			$S(p)-S_c \propto (p-p_c)^{\beta}$&$\beta$     
			&$\frac{1}{2}$~\cite{dorogovtsev2006k,lee2016critical,zhu2017revealing,li2021percolation}
			&$\frac{1}{2}$~\cite{parshani2010interdependent,gross2020interconnections}\cr
			
			
			$ \chi \propto (p-p_c)^{-\gamma} $&$\gamma$ &~1~\cite{lee2016critical}
			&~1~\cite{lee2016hybrid,dong2018resilience}\cr
			
			$S_c(N)-S_c(\infty) \propto N^{-\frac{\beta}{\bar{\nu}}}$ $^a$&$\bar{\nu}$ 
			&2~\cite{lee2016critical,zhu2017revealing}&2~\cite{lee2016hybrid}\cr
			
			$N'\propto(p-p_c)^{-\widetilde\nu}$&$\widetilde\nu$       
			&2 (Eq.~(\ref{eq5})) &2~\cite{gross2022fractal}\cr
			
			$\sigma(M_c) \propto N^{\widetilde{d}_f}$&$\widetilde{d}_f$        
			&$\frac{3}{4}$ (Eq.~(\ref{sigma_Mc})) &$\frac{3}{4}$~\cite{gross2022fractal}\cr
			
			$\sigma(M) \propto (p-p_c)^{-\epsilon }$& $\epsilon $     
			&$\frac{1}{2}$ (Eq.~(\ref{case1})) &$\frac{1}{2}$~\cite{gross2022fractal}\cr
			
			$\left\langle \tau_c\right\rangle \propto N^{\psi}$&$\psi$ 
			&$\frac{1}{3}$ (\cite{lee2016critical}, Eq.~(\ref{ave_tauc}))  &$\frac{1}{3}$~\cite{zhou2014simultaneous,lee2016hybrid}\cr
			
			$\left\langle \tau\right\rangle \propto (p-p_c)^{-\phi}$&$\phi$ 
			&$\frac{1}{2}$ (Eq.~(\ref{case2}))  &$\frac{1}{2}$~\cite{zhou2014simultaneous}\cr	
			\hline
			
		\end{tabular}\\
	$^{a}$ $S_c(N)$ denotes the fraction of the $k$-core giant component out of the original network of size $N$ at $p_c$ and $S_c(\infty)$ is for $N\to \infty$.
\end{table}
\normalsize

\section{Fractal fluctuations dimension and correlation \emph{size} of $k$-core percolation}\label{s3}
In this Section,
we study the critical exponents of the relation between the fluctuations of $p_c$ and size $N$, 
which also represents how the correlation \emph{size} scales with $p-p_c$. We further study how the fluctuations of $M$ scale with $N$ and $p-p_c$, i.e., what is the fractal fluctuations dimension. 

\subsection{Critical exponent of  fluctuations of  the critical threshold: correlation size }\label{s3.1}
\begin{figure}[htp]
	\centering
	\includegraphics[width=0.67\textwidth]{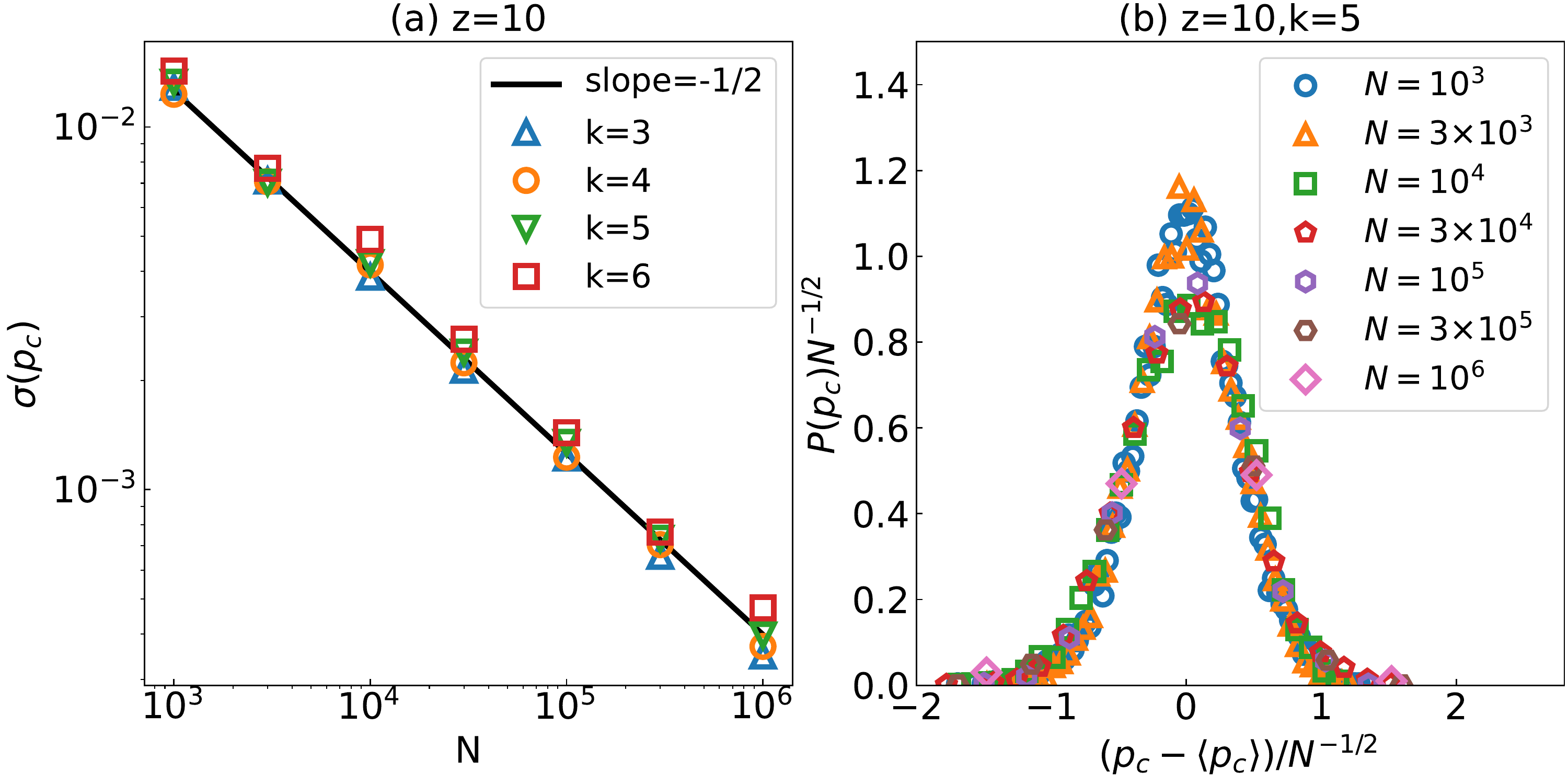}
	\caption{The correlation \emph{size} exponent of $k$-core percolation in ER network. (a) The scaling relation between the standard deviation of the $k$-core percolation threshold, $p_c$, and network size $N$.   
		It exhibits the same scaling relation as found for interdependent networks~\cite{gross2022fractal}. 
		(b) The distribution of $p_c$ follows the scaling relation of Eq. (\ref{distribution_pc}). Here we simulated $10^4$ realizations for $N\leq 3 \cdot 10^{5}$  and 200 realizations for $N=10^{6}$. Note that, we consider $k=5$ as an example for the distribution shown here, but the same holds true for other values of $k\geq3$. We have included the results for $k=3$ in the appendix, which can be seen in Fig.~\ref{figA1}(a).}
	\label{fig2}
\end{figure}

For a given network of size $N$ we study in Fig. \ref{fig2}(a) the fluctuations of $p_c$, $\sigma (p_c)$, for different realizations scale  as a function of $N$. The results suggest the following scaling with $N$, 
\begin{eqnarray}
	\sigma(p_c) \propto N^{-1/\widetilde\nu},~ \widetilde\nu=2.
	\label{sigma_pc} 
\end{eqnarray}
Thus, Eq. (\ref{sigma_pc}) suggests that 
\begin{eqnarray}
	N'\propto(p-p_c)^{-2},
	\label{eq5} 
\end{eqnarray}
is the correlation \emph{size} below which (for $N<N'$) the critical features exist and could be observed while above $N'$ the critical regime disappears (see Fig. \ref{fig3}). Note that for lattices $\sigma(p_c)$ is measured as a function of $L$, the size of the lattice \cite{gross2022fractal}, i.e., $	\sigma(p_c) \propto L^{-1/\nu'}$, suggesting correlation \emph{length} $	\xi'\propto(p-p_c)^{-\nu'}$.  Here, in ER a linear length $L$ does not exist, therefore we measure  the fluctuations as a function of $N$, satisfying $N=L^d$, where $d$ is the dimension of the spatial network. Thus, $\widetilde\nu$ can be regarded as a correlation \emph{size} exponent, 	 in analogy to the correlation \emph{length} exponent $\nu'$ in $d$-dimensional spatial network, satisfying:
\begin{eqnarray}
	\widetilde\nu \equiv d \cdot \nu'.
	\label{relation nu} 
\end{eqnarray}
In this study, we utilize finite-size scaling to measure it by following a similar  approach given in \cite{stauffer2018introduction,levinshtein1975relation,bunde2012fractals}. 

The distribution of $p_c$ for different realizations  with respect to mean $\left\langle p_c\right\rangle $ for different $N$ values is shown to behave in Fig. \ref{fig2}(b) as a scaling function,
\begin{eqnarray}
	P(p_c)N^{-1/\widetilde\nu} \sim F\left[\left(p_c-\left\langle p_c\right\rangle  \right)/ N^{-1/\widetilde\nu} \right], \label{distribution_pc}  
\end{eqnarray}
where $F(x)$ can be well approximated by a Gaussian distribution. 
Here, all distribution curves collapse into a single curve by rescaling with the correlation \emph{size} exponent $\widetilde\nu$.
Note that the obtained exponent $\widetilde\nu$ as well as the scaling in Eq.~(\ref{sigma_pc}) for k-core is the same as those found in the interdependent networks~\cite{gross2022fractal}.

\subsection{Fractal fluctuations dimension of the k-core giant component}\label{s3.2}

\begin{figure}[htp]
	\centering
	\includegraphics[width=0.67\columnwidth]{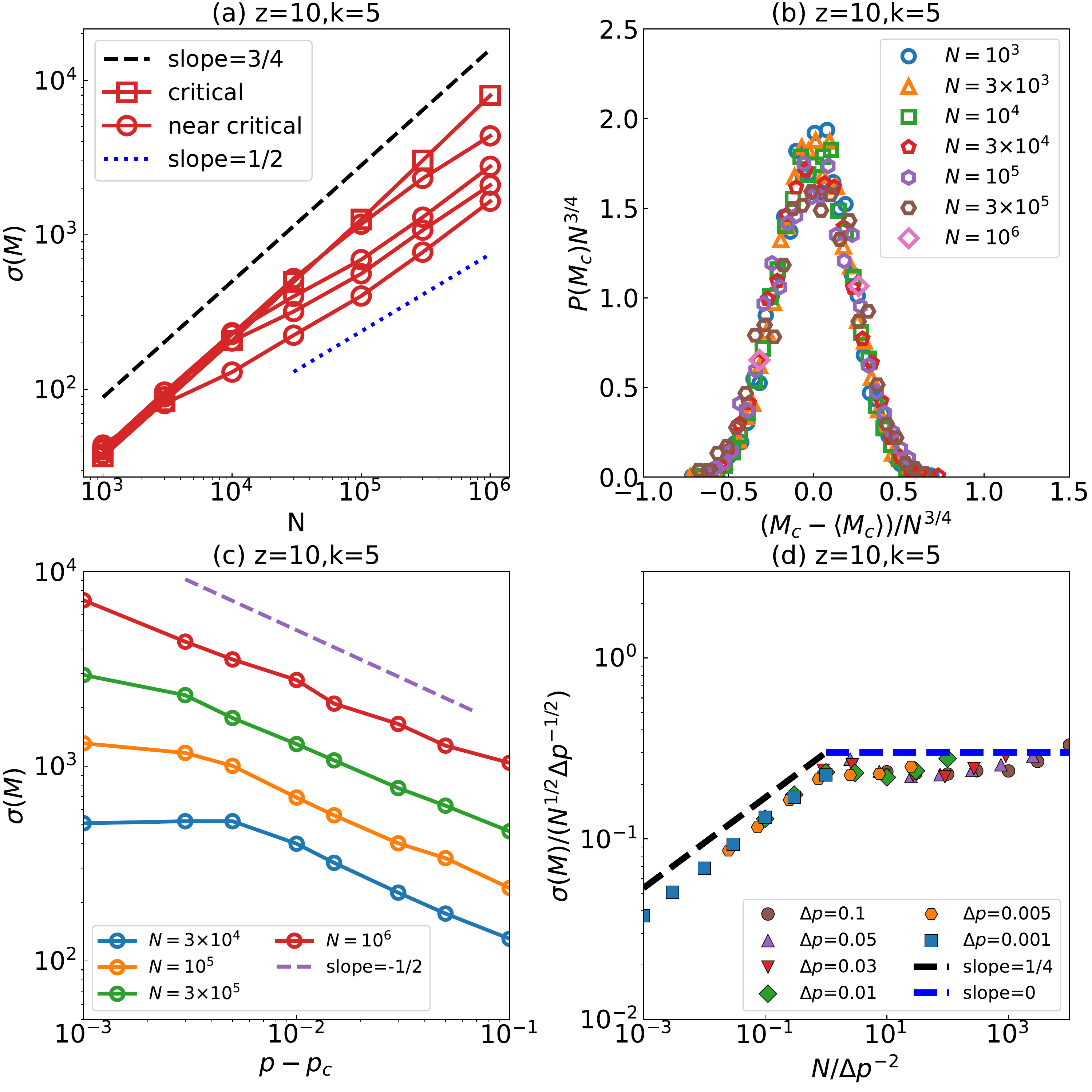}
	\caption{Fractal fluctuations dimension  of the critical mass $M_c$ of $k$-core percolation in ER network near $p_c$. 
		(a) Scaling relation between the fluctuations of $M$ and the network size $N$ at and near the  critical threshold $p_c$. Here we choose $\Delta p=p-p_c$ shown up to down being $\left\lbrace 0.003,0.01,0.015,0.03\right\rbrace $. The crossover in the slope represented by the exponent $\widetilde{d}_f$ is  clearly seen. That is, within a correlation \emph{size}, $N'$ (Eq. (\ref{eq5})), fractality can be observed with $\widetilde{d}_f=3/4$ while above the correlation \emph{size}, there is no fractality and Eq. (\ref{eq8}) appears.
		(b) The distribution of $M_c$ follows the scaling relation in Eq. (\ref{distribution_MC}). 
		(c) Scaling relation between the fluctuations of $M$ and $p-p_c$. Note that, the scaling relation between the fluctuations of $S$ and $p-p_c$ (see Eq.~(\ref{gamma_S})) has been also investigated in  \cite{lee2016critical}. $\sigma(M)\propto (p-p_c)^{-1/2}$ can be derived from  Eq.~(\ref{gamma_S}), further supporting our findings here.
		(d) Fractal fluctuations behavior in the variation of the mass $M$ near the critical mixed-order percolation threshold, supporting Eq. (\ref{case1}). Our results for k-core follow the same scaling function, crossover and critical exponents found in interdependent networks~\cite{gross2022fractal}. Note that, an analogous crossover has been also observed in  \cite{lee2016critical}, further supporting our fractality findings here. We considered  here $k=5$ as an example, but the same holds true for other values of $k\geq3$ as can be seen in the appendix in  Fig.~\ref{figA1}(b) and Fig.~\ref{figA2}.}
	\label{fig3}
\end{figure}

For a given $N$ we study in Fig. \ref{fig3}(a) the fluctuations of the k-core giant component $M$, $\sigma (M)$, at and near the threshold $p_c$ based on different realizations, and then plot it as a function of $N$. The results at criticality suggest the following scaling with $N$, 
\begin{eqnarray}
	&\sigma(M_c) \propto N^{\widetilde{d}_f} ,~\widetilde{d}_f=3/4
	\label{sigma_Mc}.
\end{eqnarray}
Here $\widetilde d_f$ is the fractal fluctuations dimension
of the k-core giant component in random networks with respect to the size $N$. While in $d$-dimensional lattices,  $\sigma(M_c)$ is measured as a function of $L$, i.e., $\sigma(M_c) \propto L^{d'_f} \cite{gross2022fractal}$. Since $N=L^d$, we have
\begin{eqnarray}
	\widetilde d_f \equiv  d'_f/d.
	\label{relation df} 
\end{eqnarray}
Further, similar to the previous section, by rescaling with the fractal fluctuations dimension, $\widetilde{d}_f$, we collapse in Fig. \ref{fig3}(b), the distribution of $M_c$ for different $N$ according to the scaling, 
\begin{eqnarray}
	P(M_c)N^{3/4} \sim F\left[\left(M_c-\left\langle M_c\right\rangle  \right)/ N^{3/4} \right]. 
	\label{distribution_MC} 
\end{eqnarray}
While Eq.~(\ref{sigma_Mc}) is valid at the critical threshold, away from the critical threshold and above the correlation \emph{size}, we can see the normal scaling, i.e., 
\begin{eqnarray}
	\sigma (M) \sim N^{1/2}. 
	\label{eq8} 
\end{eqnarray}	
Close to the critical threshold, a crossover between these two behaviors is observed, which can be described using a scaling function $f(\mu)$ as	
\begin{eqnarray}
	\sigma(M) \propto N^{3/4}f(\mu) \label{eq4} \label{sigma_M},
\end{eqnarray}
with
\begin{eqnarray}
	\mu=(p-p_c)^{\alpha}\cdot N,
\end{eqnarray}
where $f(\mu)$ is a piecewise function satisfying $f(\mu) \propto$ constant for $\mu<1$ and $f(\mu) \propto \mu^{m}=(p-p_c)^{\alpha m}\cdot N^m$ for $\mu>1$. Thus, we have:
\begin{equation}
	\label{cases}
	\sigma(M)\propto\cases{N^{3/4}&for $\mu<1$\\
		N^{3/4+m}(p-p_c)^{\alpha m}&for $\mu>1$\\}.
\end{equation}
In Fig.~\ref{fig3}(a) we can see that $\sigma(M) \propto N^{1/2}$ for $\mu>1$, implying  $3/4+m=1/2$, i.e., $m=-1/4$. Thus, we have:
\begin{equation}
	\label{cases}
	\sigma(M)\propto\cases{N^{3/4}&for $\mu<1$\\
		N^{1/2}(p-p_c)^{-\alpha/4}&for $\mu>1$\\}.
\end{equation}
In order to determine the value of $\alpha$, we plot $\sigma(M)$ against $p-p_c$ in Fig.~\ref{fig3}(c). 
We obtain  $\sigma(M)\propto (p-p_c)^{-1/2}$, implying that $\alpha=2$. 
Finally, 
to support Eq. (\ref{sigma_M}) and the obtained value of $\alpha$, 
we created a scaled plot shown in Fig.~\ref{fig3}(d), 
depicting $\sigma(M)/\left( N^{1/2}(\Delta p)^{-1/2}\right) $ against $N(\Delta p)^2$.
As can be seen, 
we achieve a satisfactory scaling collapse with $\alpha=2$, i.e., we have:
\begin{equation}
	\label{case1}
	\sigma(M)\propto\cases{N^{3/4}&for $N<N'\propto (p-p_c)^{2}$\\
		N^{1/2}(p-p_c)^{-1/2}&for $N>N'$\\}.
\end{equation}
Thus, k-core features again the same universal scaling function as that of interdependent network~\cite{gross2022fractal}.

\section{Hyper-scaling relation for fluctuations in mixed-order transition}\label{s4}
In this Section, we establish analytical arguments for a universal hyper-scaling relation for the fractal fluctuations dimension of the order parameter in a mixed-order transition in spatial networks of size, $N=L^{d}$, which also provides insight for non-spatial random networks. 

We start with the scaling of the  fraction of $k$-core giant component, $S$, close to criticality for both second-order and mixed-order transitions,
\begin{eqnarray}
	S(p)-S_{c}\propto (p-p_c)^{\beta}. \label{beta} 
\end{eqnarray}
Here, for second order transition $S_c=0$ while for mixed order $S_c$ is finite.
Next, we substitute the size (number of sites) of the giant component, $M_c$, at $p_c$ and $M(p)$ away (but closeby) from the critical threshold
\begin{eqnarray}
	\frac{M(p)}{L^d}-\frac{M_c}{L^d}\propto (p-p_c)^{\beta}. \label{beta_mass} 
\end{eqnarray}
Based on the general case, Eq. (\ref{beta_mass}), 
one can obtain the well-known hyperscaling relation 
for continuous second-order transitions~\cite{stauffer2018introduction,bunde2012fractals}. 
Substituting $M_c=0$ and $M(p) \propto L^{d_f}$ into Eq.~(\ref{beta_mass}), 
yield $L^{d_f}/L^d\propto(p-p_c)^\beta$ and since $\xi \propto(p-p_c)^{-\nu}$, the hyperscaling relation
\begin{eqnarray}
	d_{f}=d-\beta/ \nu \label{d_f} 
\end{eqnarray}
is derived.

Different from the continuous phase transition case, in the mixed-order phase transition case, for Eq. (\ref{beta_mass}), we evaluate instead the standard deviation of both sides
\begin{eqnarray}
	\sigma\left( \frac{M(p)}{L^d}-\frac{M_c}{L^d}\right) \propto \sigma\left( (p-p_c)^{\beta}\right). \label{sigma_delta} 
\end{eqnarray}
Using the known relation $\sigma(X^a) \sim X^{a-1} \sigma (X)$ \cite{ku1966notes} and assuming that the dominant variations between different  realizations are in $p_c$ and $M_c$, we get
\begin{eqnarray}
	\frac{	\sigma\left( M_c\right) }{L^d}\propto(p-p_c)^{\beta-1} \sigma\left( p-p_c\right). \label{sigma_delta_eq1} 
\end{eqnarray}
	Assuming,  $\sigma \left( M_c \right) \propto L^{d'_{f}}$, for $L<\xi'$, we get 
	\begin{eqnarray}
		\frac{	L^{d'_{f}}}{L^d}\propto(p-p_c)^{\beta-1} \sigma\left( p_c\right). \label{sigma_delta_eq2} 
	\end{eqnarray}
	Substituting  into Eq. (\ref{sigma_delta_eq2}) the scaling $\xi'\propto(p-p_c)^{-\nu'}$ and assuming that the variation of $p_c$ are controlled by $\xi'$, we obtain,
	\begin{eqnarray}
		L^{d'_{f}-d}=\xi'^{-\frac{\beta-1}{\nu'}} \xi'^{-\frac{1}{\nu'}}. \label{Eq with xi 1} 
	\end{eqnarray}
	Taking the limit $L \to  \xi' $, we obtain,
	\begin{eqnarray}
		\xi'^{d'_{f}-d}=\xi'^{-\frac{\beta}{\nu'}}.  \label{Eq with xi 2} 
	\end{eqnarray}	
	Thus, deriving the new but analogous to Eq. (\ref{d_f}) hyperscaling relation, Eq.~(\ref{hyper}), for the fractal fluctuations dimension, $d'_f=d-\beta\nu'$.
	
	Note that the hyperscaling relations given by Eq.~(\ref{hyper}) (for mixed-order phase transitions) and Eq.~(\ref{d_f}) (for second-order phase transitions) are similar, 
	but the interpretation of the fractal dimensions $d'_f$ and $d_f$ are different. For second-order continuous phase transitions, $d_f$ describes the fractal dimension of the bulk~\cite{stauffer2018introduction,bunde2012fractals}, while for mixed-order phase transitions, $d'_f$ describes the fractal dimension of the fluctuations of $M_c$, i.e., how $\sigma(M_c)$, scales with $L$. Below the correlation \emph{length} the fluctuations  are fractals, but above they behave normally. Note that the hyperscaling relation, Eq. (\ref{hyper}) holds for spatial networks of different dimension $d$. Correspondingly, it also holds for random networks like the ER networks. 
	Substituting Eq. (\ref{relation nu}) and Eq. (\ref{relation df}) into Eq. (\ref{hyper}), we can obtain the hyperscaling relation for random networks, 
	\begin{eqnarray}
		\widetilde{d}_f=1-\beta/\widetilde \nu. \label{hyperscaling relation_ER} 
	\end{eqnarray}	
	As expected, the evaluated critical exponents $\beta=1/2$, $\widetilde\nu=2$ and $\widetilde{d}_f=3/4$ found here satisfy this hyperscaling relation, Eq. (\ref{hyperscaling relation_ER}). Noteworthy, the hyperscaling relation Eq. (\ref{hyper}) with identical exponents as found here for k-core, is also valid for interdependent networks~\cite{gross2022fractal}.
	
	\section{Discussions and Summary}\label{s5}
	In the present study, we investigate systematically the $k$-core percolation phase transitions in ER random networks to uncover the similarities of percolation in  the k-core  and interdependent networks models, along with identifying the possible underlying mechanisms for the similarities.
	We examine several new critical exponents found recently in IN \cite{gross2022fractal}, such as the critical exponents of fluctuations of $p_c$, representing the diverging correlation \emph{size}, $N'$, and of the order parameters in different realizations of size, $N$, representing the fluctuation fractal dimension. 
	The findings are summarized in Table 1, where the common critical exponents of $k$-core percolation and interdependent networks reveal intriguing similarities between the two models. 
	
	Our results  suggest that the similarity of the criticality between the two systems (and probably to many other systems) originates from the two types of interactions that exist in both systems, one is the connectivity links which are short range and the other is the $k$-core $(k\geq3)$ or the dependency (large $\ell$) which are both long-range.
	When plotting the distributions of distances 
	caused by a failure in different cores,
	we found that for $k\geq 3$ the interactions are LR, while for $k=1$ and $k=2$ the interactions are SR (see Fig. \ref{fig1_kcore}).
	Similar to $k$-core percolation, percolation results of the CD model again show that SR interactions exist for $\ell=0$ and $\ell=1$ and thus yield a second order phase transition, 
	while LR interactions, e.g., $\ell=D$ yields a mixed order phase transition. 
	In both models (and potentially also for other models) the mechanisms behind the different phase transitions seem the same, and short-range and long-range influences are the origin.
	Moreover, we demonstrate that
	the fractal fluctuations dimension of the order parameter in the mixed-order phase transition satisfies the hyperscaling relation, $d'_{f}=d-\beta/\nu'$. 
	Our study offers valuable insights into the striking similarities   between $k$-core percolation and percolation in interdependent networks, shedding light on the phenomenon of fractal fluctuations dimension observed in mixed-order phase transitions.

	\ack
	We thank the Israel Science Foundation, the Binational Israel-China Science Foundation Grant No.\ 3132/19, NSF-BSF Grant No.\ 2019740, the EU H2020 project RISE (Project No. 821115), and the EU H2020 DIT4TRAM for financial support. 
	Li D acknowledges the support from National Natural Science Foundation of China (72225012, 71822101, 71890973/71890970). She Z  acknowledges the support from the National Key R$\&$D Program of China (No.2022YFA1005103).
	Gross B acknowledges the support of the Mordecai and Monique Katz Graduate Fellowship Program. Gao S and Xue L acknowledge the support from the program of China Scholarship Council.

	\section*{References}
	%
	%
	%
	%
	\bibliographystyle{iopart-num}
	\bibliography{k_core_bib}

	\appendix
	\section{Critical exponent $\beta$ in CD model for $\ell=D$}\label{A4}
	Here we consider the single-layer network model that incorporates both connectivity links and dependency links, where the dependency links are subject to a maximal shortest path length constraint $\ell$ called here the CD model, see Fig.~\ref{fig1_dependency}(a). When $\ell=D$, where D is the network diameter, dependency pairs are randomly selected. Parshani et al. \cite{parshani2011critical} derived the formula:
	\begin{eqnarray}
		&S(p)=p^2\left( 1-\exp(-zS(p))\right)^2   \label{theo}. 
	\end{eqnarray}	
	suggesting:
	\begin{eqnarray}
		&\exp(-zS)=1-S^{1/2}/p   \label{relation}. 
	\end{eqnarray}	
	Here for simplicity, we denote $S(p)$ as $S$.
	To analytically find the critical exponent $\beta$, we define:
	\begin{eqnarray}
		&f(p,S)=S-p^2\left( 1-\exp(-zS)\right)^2   \label{function}. 
	\end{eqnarray}
	Letting $f(p,S)=0$, Eq. (\ref{theo}) can be obtained. 
	To find the critical exponent $\beta$ near the critical point $(p_c,~S_c)$, we can expand $f(p,S)$ near $p=p_c$ and $S=S_c$:
	\begin{eqnarray}
		&f(p,S)=f(p_c,S_c)+f'_S(p_c,S_c)(S-S_c)+f'_p(p_c,S_c)(p-p_c) \nonumber
		\\ 
		&+\frac{f''_{SS}(p_c,S_c)}{2!}(S-S_c)^2+\frac{f''_{pp}(p_c,S_c)}{2!}(p-p_c)^2
		\nonumber
		\\ 
		&+\frac{f''_{Sp}(p_c,S_c)+f''_{pS}(p_c,S_c)}{2!}(S-S_c)(p-p_c)+...=0
		\label{expand function}.
	\end{eqnarray}
	Here, $f'_S$ and $f'_p$ denotes the  partial derivatives of $f$ with respect to $S$ and $p$;  $f''_{SS}$ and $f''_{pp}$ denotes the second order partial derivatives of $f$ with respect to $S$ and $p$; $f''_{Sp}$ and $f''_{pS}$ denotes the second mixed partial derivatives of $f$.
	At the critical point $(p_c,~S_c)$,
	calculating $f'_S$, $f'_p$ and $f''_{SS}$ and substituting  into Eq~(\ref{relation}), we obtain
	\begin{eqnarray}
		&f'_S(p_c,S_c)=2zS_c-2zp_cS_c^{1/2} + 1=0 \label{coefficient1}
		\\ 
		&f'_p(p_c,S_c)=-2S_c/p_c\neq 0 \label{coefficient2}
		\\ 
		&f''_{SS}(p_c,S_c)=-zp_cS_c^{-1/2}+2z\neq 0
		\label{coefficient3}.
	\end{eqnarray}
	Here, $f'_S(p_c,S_c)=0$ is the condition for the first-order abrupt phase transition \cite{parshani2011critical}, and we know that  Eq. (\ref{theo}) yields first-order abrupt phase transition. $f'_p(p_c,S_c)\neq 0$ is because $S_c\neq 0$. $f''_{SS}(p_c,S_c)\neq 0$ holds, since if it is not true, it will lead to a contradiction with Eq.~(\ref{coefficient1}). 
	Then rearranging Eq.~(\ref{expand function}) yields
	\begin{eqnarray}
		&(S-S_c)^2=-2\frac{f'_p(p_c,S_c)}{f''_{SS}(p_c,S_c)}(p-p_c)\nonumber
		\\ 
		&-2\frac{f''_{Sp}(p_c,S_c)+f''_{pS}(p_c,S_c)}{f''_{SS}(p_c,S_c)}(S-S_c)(p-p_c)+...
		\label{scaling relation}
	\end{eqnarray}
	Here, since the first term in Eq.~(\ref{scaling relation}) is the leading term, we derive the scaling relation $S-S_c\propto (p-p_c)^{1/2}$, i.e., $\beta=1/2$, supporting Fig.~\ref{fig1_dependency}(d).

	\section{Critical exponent in $k$-core percolation}
	\subsection{The collapsed Probability Density Function (PDF) in $k$-core percolation}\label{A1}
\begin{figure}[htp]
		\centering
		\includegraphics[width=0.67\columnwidth]{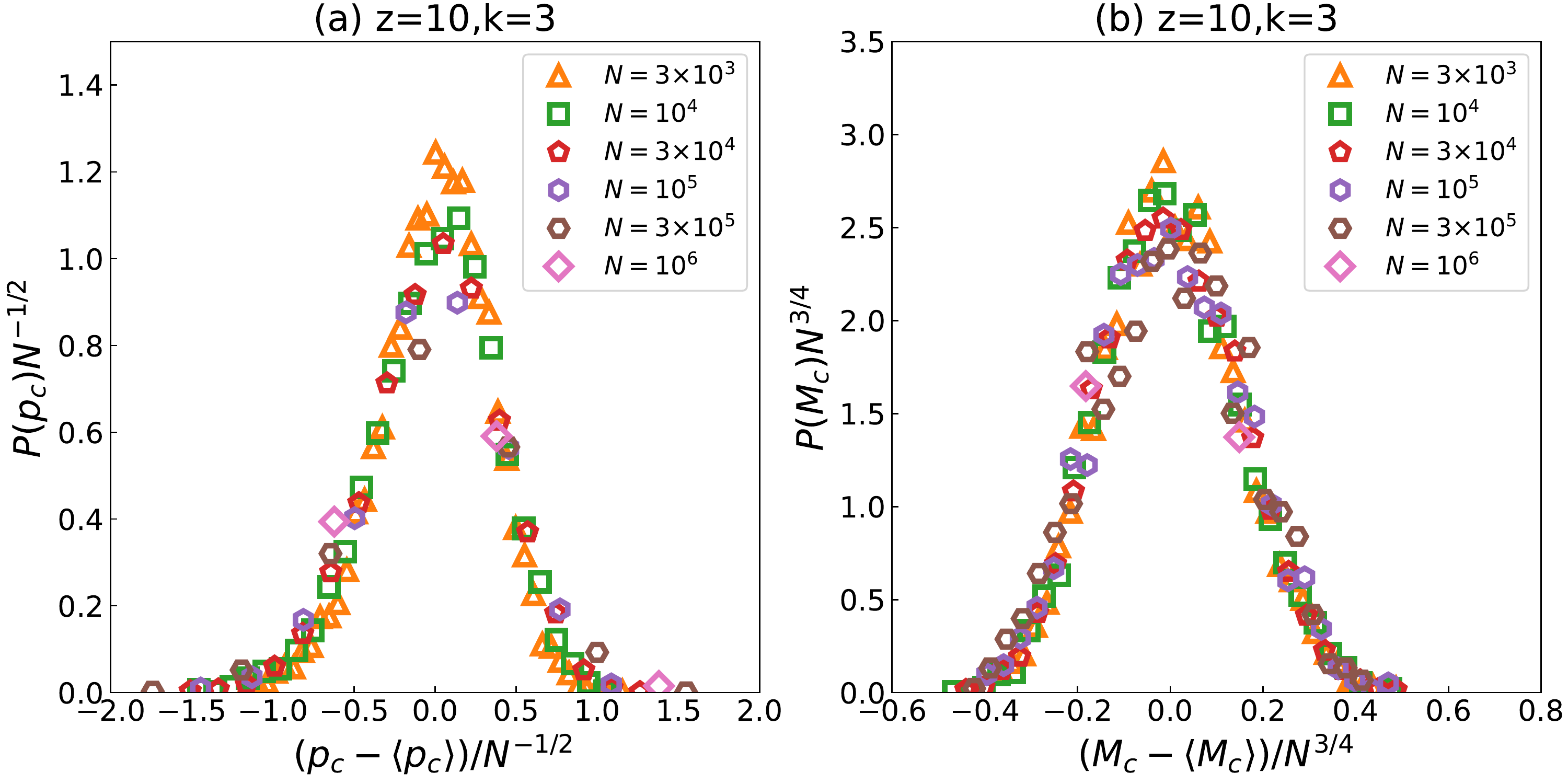}
		\caption{The collapsed Probability Density Function (PDF). 
			(a) The distribution of $p_c$ follows the scaling relation in Eq. (\ref{distribution_pc}). 
			(b) The distribution of $M_c$ follows the scaling relation in Eq. (\ref{distribution_MC}). 
			It exhibits the same scaling relation as in interdependent networks~\cite{gross2022fractal}.}
		\label{figA1}
	\end{figure}

	Here for $k=3$ core percolation in ER network, we collapse the PDF of critical threshold $p_c$ (critical mass $M_c$) together utilizing the same scaling relationship given in Eq.~(\ref{sigma_pc}) (Eq.~(\ref{distribution_MC})), as shown in Fig.~\ref{figA1}. 
	
	\subsection{Fractal fluctuations of the k-core giant component}\label{A2}
\begin{figure}[htp]
		\centering
		\includegraphics[width=\columnwidth]{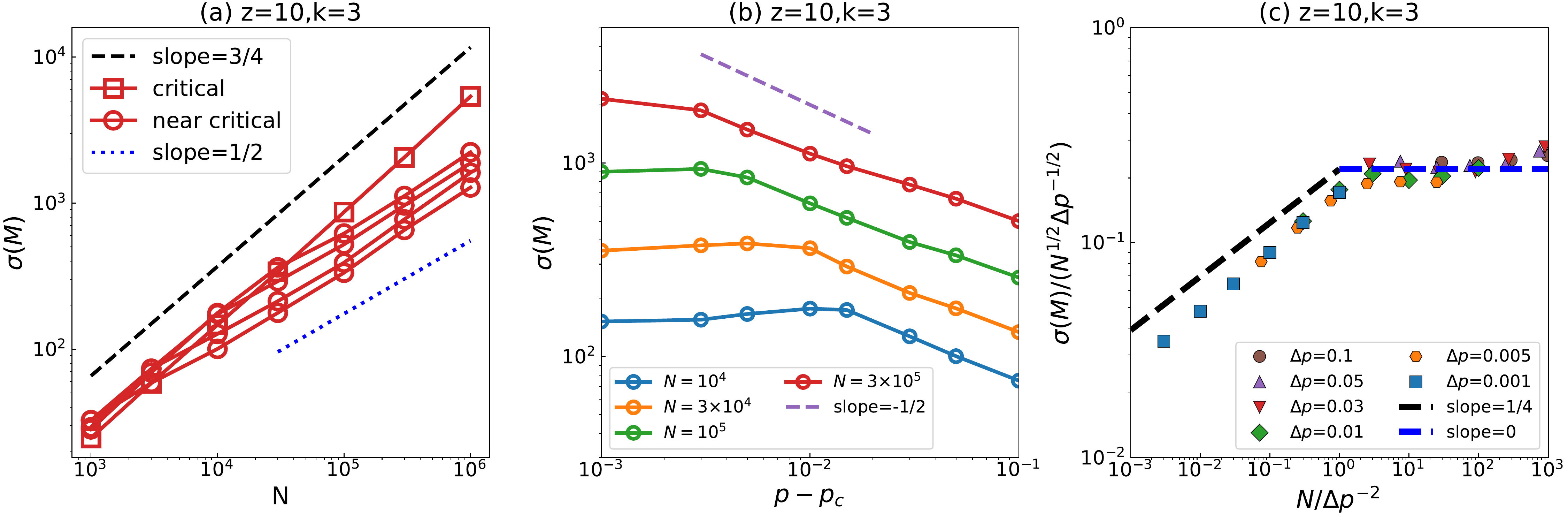}
		\caption{Critical exponents of fluctuations of the critical mass $M_c$ of $k$-core percolation in ER network near $p_c$. 
			(a) Scaling relation between the fluctuations of $M$ and the network size $N$ at and near the  critical threshold $p_c$. Here we choose $\Delta p=p-p_c$ from up to down to be $\left\lbrace 0.01,0.015,0.03,0.05\right\rbrace $. The crossover in the exponent $\widetilde{d}_f$ is  clearly seen. 
			(b) Scaling relation between the fluctuations of $M$ and $p-p_c$. (c) Fractal fluctuation behavior in the standard deviation of the mass $M$ near the critical mixed-order percolation threshold, supporting Eq. (\ref{case1}). It follows the same scaling function, crossover and critical exponents as found in interdependent networks~\cite{gross2022fractal}. }
		\label{figA2}
	\end{figure}

	For $k=3$ core percolation in ER network, the scaling behavior of  fluctuation of $k$-core giant component at and near threshold is plotted in Fig.~\ref{figA2}, which shows the same critical behavior as for k=5  shown in Fig~\ref{fig2}, Subsection \ref{s3.2}. Note that it is the same scaling function found for percolation of interdependent networks with the same exponents~\cite{gross2022fractal}.
	
	\subsection{Scaling behaviors of mean plateau time in $k$-core percolation}\label{s3.3}
\begin{figure}[htp]
		\centering
		\includegraphics[width=0.67\columnwidth]{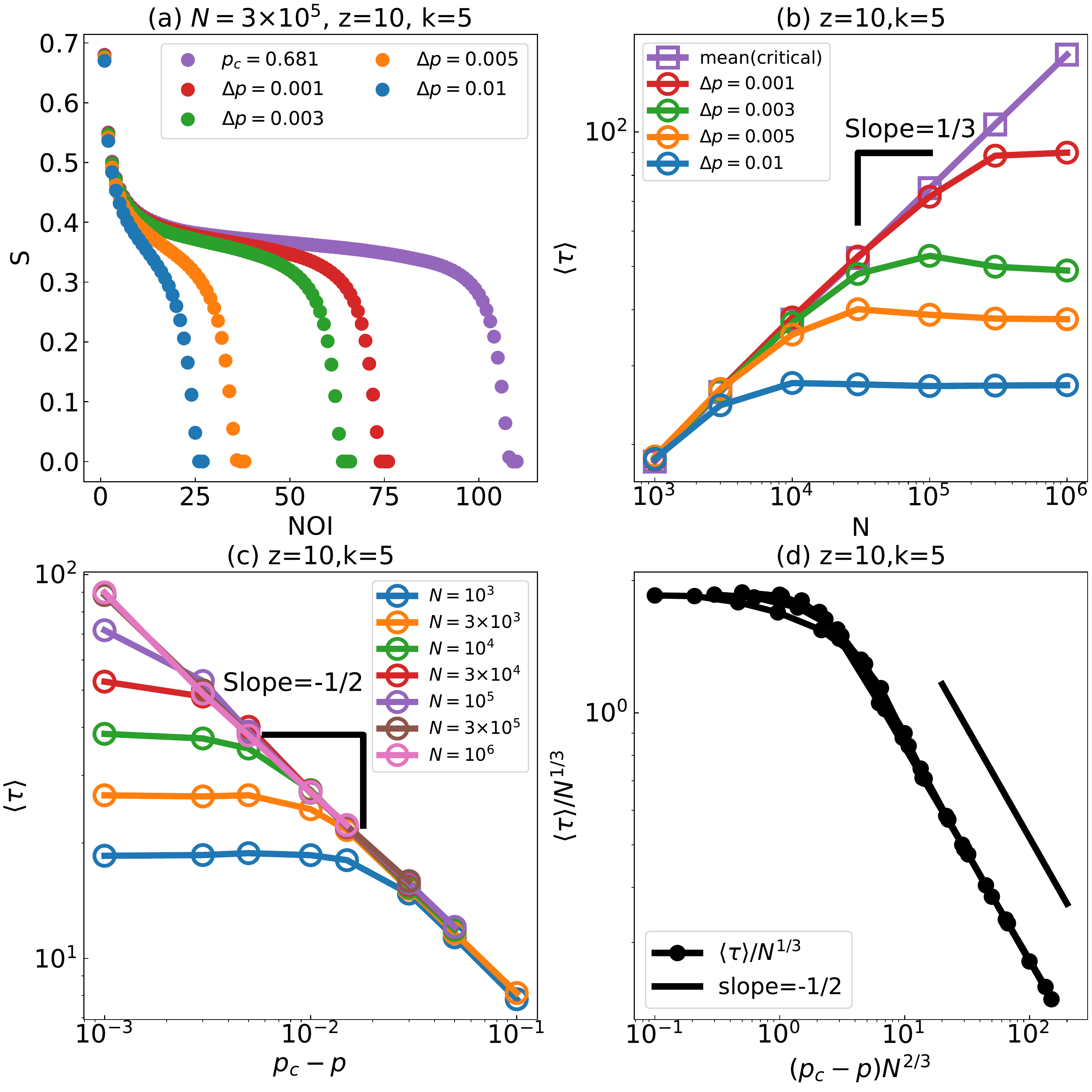}
		\caption{The plateau  of single realizations. 
			(a) Dynamical process, i.e., the plateau behavior of the fraction of $k$-core giant component, $S$, of a network with mean degree $z$ and size $N$ with NOI (time) of $k$-core subgraph at criticality, $p_c$ and just below $p_c$. 
			(b) Scaling relations between the mean plateau time (number of iterations-$\tau$) and the network size $N$ at critical $p_c$ and at four non-critical $p$ values just below $p_c$. Note that, the relation $\left\langle \tau_c \right\rangle  \propto N^{1/3}$ was also found in  \cite{lee2016critical}, supporting the findings here.
			(c) Scaling relations of the mean plateau time, $\left\langle \tau\right\rangle $,  against $p_c-p$ for different $N$. 
			(d) The scaling collapse of (b) and (c), supports Eq. (\ref{scaling tao}). It is seen that the scaling behavior of the mean plateau time and their critical exponents for $k$-core are identical to those of interdependent networks~\cite{zhou2014simultaneous}. }
	\label{fig4}
\end{figure}

	For a given $N$ we study in Fig. \ref{fig4}(a) the NOI which we call the mean plateau time $\left\langle \tau\right\rangle $ at and near the threshold $p_c$ for different realizations, and then plot it in Fig. \ref{fig4}(b) as a function of $N$. The results suggest the following scaling with $N$, 
\begin{eqnarray}
	&\left\langle \tau_c \right\rangle  \propto N^{1/3} \label{ave_tauc},
\end{eqnarray}
in agreement with Lee et al. \cite{lee2016critical}. While Eq.~(\ref{ave_tauc}) is valid at the critical $p_c$, away from the critical regime fluctuations are observed in Fig. \ref{fig4}(b) independent of $N$, i.e., as $\sim N^{0}$. Close to the critical threshold, a crossover between these two behaviors is observed and can be described via a scaling function $f_1(\mu_1)$ as	
\begin{eqnarray}
	\left\langle \tau \right\rangle  \propto N^{1/3}f_1(\mu_1) \label{eq4} \label{mean_tau },
\end{eqnarray}
with
\begin{eqnarray}
	\mu_1=(p_c-p)^{\alpha_1} \cdot N.
\end{eqnarray}
Here, $f_1(\mu_1)$ is a piecewise function satisfying $f_1(\mu_1) \propto$ constant for $\mu_1<1$ and $f_1(\mu_1) \propto \mu^{m_1}=(p_c-p)^{\alpha_1 m_1} \cdot N^{m_1}$ for $\mu_1>1$. Thus, we have:
\begin{equation}
	\label{cases}
	\left\langle \tau \right\rangle \propto\cases{N^{1/3}&for $\mu_1<1$\\
		N^{1/3+m_1}(p_c-p)^{\alpha_1 m_1}&for $\mu_1>1$\\}.
\end{equation}
From Fig.~\ref{fig4}(b) we can see that $\left\langle \tau_c \right\rangle  \propto $ constant  for $\mu>1$, implying  $1/3+m_1=0$, i.e., $m_1=-1/3$. Thus, we have:
\begin{equation}
	\label{scaling tao}
	\left\langle \tau \right\rangle \propto\cases{N^{1/3}&for $\mu_1<1$\\
		(p_c-p)^{-\alpha_1/3}&for $\mu_1>1$\\}.
\end{equation}
In order to determine the value of $\alpha_1$, 
we plot $\left\langle \tau_c \right\rangle $ against $p_c-p$ in Fig.~\ref{fig4}(c). 
We can see that $\left\langle \tau_c \right\rangle \propto (p_c-p)^{-1/2}$, 
implying $\alpha_1=3/2$. 
Finally, 
to support Eq. (\ref{mean_tau }) and the obtained value of $\alpha_1$, 
we created a scaled plot shown in Fig.~\ref{fig3}(d), 
depicting $\left\langle \tau_c \right\rangle / N^{1/3} $ against $N^{2/3}\Delta p$. 
As can be seen, 
one achieves a satisfactory scaling collapse with $\alpha_1=3/2$, i.e., we have:
\begin{equation}
	\label{case2}
	\left\langle \tau \right\rangle \propto\cases{N^{1/3}&for $N<N_1'\propto(p_c-p)^{-3/2}$\\
		(p_c-p)^{-1/2}&for $N>N_1'$\\}.
\end{equation}
showing further that the scaling behavior of the mean plateau time and its critical exponents in $k$-core subgraphs are identical to those of interdependent networks~\cite{zhou2014simultaneous}.

%
\end{document}